\documentclass[]{aa}
\usepackage{graphics}
\begin{document}
\title{High-resolution \ion{O}{vi} absorption line observations at
$1.2 \leq z \leq 1.7$ in the bright QSO HE 0515-4414
\thanks{Based on observations with the NASA/ESA Hubble Space
Telescope, obtained at the Space Telescope Science Institute,
which is operated by Aura, Inc., under NASA contract NAS 5-26555;
and on observations collected at the VLT/Kueyen telescope, ESO,
Paranal, Chile.}}
\author {Dieter Reimers \inst{1}, Robert Baade \inst{1}, Hans-J\"urgen Hagen
\inst{1}, and Sebastian Lopez \inst{2}} \offprints{D.~Reimers}
\institute{Hamburger Sternwarte, Universit\"at Hamburg,
Gojenbergsweg 112, D-21029 Hamburg, Germany \\
email: dreimers@hs.uni-hamburg.de, rbaade@hs.uni-hamburg.de, hhagen@hs.uni-hamburg.de\\
\and Departamento de Astronomia, Universidad de Chile, Casilla 36-D, Santiago, Chile\\
email: slopez@rhea.das.uchile.cl}
\date{received date; accepted date}
\titlerunning{\ion{O}{vi} absorption lines in the bright
QSO HE 0515-4414}
\authorrunning{D.~Reimers et al.}
\abstract {STIS Echelle observations at a resolution of $10\,{\rm
km\,s^{-1}}$ and UVES/VLT spectroscopy at a resolution of $7\,{\rm
km\,s^{-1}}$ of the luminous QSO HE 0515-4414 ($z_{\rm em} =
1.73$, $B = 15.0$) reveal four intervening \ion{O}{vi} absorption
systems in the redshift range $1.21 \leq z_{\rm abs} \leq 1.67$
(1.38503, 1.41601, 1.60175, 1.67359). In addition two associated
systems at $z = 1.69707$ and $z = 1.73585$ are present. Noteworthy
is an absorber at $z = 1.385$ with $\log\,N_{\rm \ion{H}{i}} =
13.9$ and strong \ion{O}{vi} (N(\ion{O}{vi})/N(\ion{H}{i})
$\approx$ 1) and \ion{C}{iv} doublets, while a nearby much
stronger Ly $\alpha$ absorber (log $N_{\rm \ion{H}{i}} = 14.8$, $\Delta v =
123\,{\rm km\,s^{-1}}$) does not reveal any heavy element
absorption. For the first time high resolution observations allow
to measure radial velocities of \ion{H}{i}, \ion{C}{iv} and
\ion{O}{vi} simultaneously in several absorption systems (1.385,
1.674, 1.697) with the result that significant velocity
differences (up to $18\,{\rm km\,s^{-1}}$) are observed between
\ion{H}{i} and \ion{O}{vi}, while smaller differences (up to
$5\,{\rm km\,s^{-1}}$) are seen between \ion{C}{iv} and
\ion{O}{vi}. We tentatively conclude that \ion{H}{i}, \ion{O}{vi},
and \ion{C}{iv} are not formed in the same volumes and that
therefore implications on ionization mechanisms are not possible
from observed column density ratios \ion{O}{vi}/\ion{H}{i} or
\ion{O}{vi}/\ion{C}{iv}. The number density of \ion{O}{vi}
absorbers with $W_{\rm rest} \geq 25\,$m\AA\ is ${\rm d}N/{\rm d}z
\leq 10$, roughly a factor of 5 less than what has been found by
Tripp at al.\ (2000) at low redshift. However, this number is
uncertain and further lines of sight will be probed in the next
HST cycle. An estimate of the cosmological mass-density of the
\ion{O}{vi}-phase yields $\Omega_{\rm b}(\ion{O}{vi}) \approx
0.0003\, h^{-1}_{75}$ for $\rm [O/H] = -1$ and an assumed
ionization fraction \ion{O}{vi}/O = 0.2. It should be noted that
this result is subject to large systematic errors. This
corresponds to an increase by roughly a factor of 15 between
$\bar{z} = 1.5$ (this work) and the value found by Tripp et al.\
(2000) at $\bar{z}$ = 0.21, if the same oxygen abundance $\rm
[O/H] = -1$ is assumed. Agreement with the simulations by Dav\'e
et al.\ (2001) can be obtained, if the oxygen abundance increases
by a factor of $\sim 3$ over the same redshift interval.
\keywords{cosmology: observations -- intergalactic medium --
quasars: absorption lines -- quasars: individual: HE 0515-4414} }
\maketitle

\section{Introduction}
Recent observations of intervening \ion{O}{vi} absorbers in
HST-STIS Echelle spectra of bright, low redshift QSOs have
provided strong evidence that in the local universe a considerable
fraction of baryonic matter might be ''hidden'' in a warm ($\sim
10^{5}\,$K) intergalactic medium (Savage et al.\ 1998; Tripp et
al.\ 2000; Tripp \& Savage 2000). This observation is in
accordance with models of hierarchical structure formation by Cen
\& Ostriker (1999) and Dav\'e et al.\ (2001) which predict that a
considerable fraction of all baryons reside in a warm-hot phase of
the intergalactic medium (WHIM) shock-heated to temperatures of
$10^{5} - 10^{7}\,$K. The same models predict that the fraction of
baryons residing in this WHIM increases strongly with decreasing
redshift from less than 5\,\% at $z = 3$ to $30 - 40$\,\% at $z =
0$. Can this prediction be verified or disproved by observations,
or can observations even impose constraints on the models? This
appears difficult for various reasons. First of all, the WHIM is
difficult to detect (cf. Dav\'e et al.\ 2001), both as diffuse
X-ray emission of the hotter parts or in absorption through the
\ion{O}{vi} doublet. In addition the temperature distribution of
the WHIM varies with redshift so that a complete census would
require the detection of all components as a function of redshift.
The warm \ion{O}{vi} component has the additional complication
that both the oxygen abundance and the ionization process cannot
be determined from \ion{O}{vi} observations alone. While at low
redshift ($z < 0.3$) collisional ionization is the most probable
process since the ionizing extragalactic UV background is diluted,
\ion{O}{vi} can be produced easily by photoionization at redshifts
$\geq 2$ and has been observed to be ubiquitous in the low-density
IGM (Schaye et al.\ 2000). On the other hand \ion{O}{vi} is not
expected to be produced by photoionization for $z \geq 3$ since
the reionization of \ion{He}{ii} is incomplete (Reimers et al.\
1997; Heap et al.\ 2000) and the IGM therefore opaque to photons
with energies above 8.4 Rydberg. Remains the intermediate redshift
range which for $z < 1.9$ requires high-resolution UV-spectroscopy
of a bright, high-redshift QSO. In this paper, we present combined
high-resolution HST/STIS observations of \ion{O}{vi} absorption
supplemented by ESO-VLT/UVES spectroscopy of the accompanying
\ion{H}{i} and \ion{C}{iv} lines in the brightest known
intermediate redshift QSO HE 0515-4414 ($z_{\rm em} = 1.73$, $B =
15.0$) discovered by the Hamburg/ESO Survey (Reimers et al.\
1998). The data have been taken mainly with the aim to study the
evolution of the Ly $\alpha$ forest and its metal content in the
range $z = 1$ to 1.7. In this first paper we concentrate on
intervening \ion{O}{vi} absorption.
\section{Observations and data reduction}
\subsection{Hubble Space Telescope observations}
HE 0515-4414 was observed with STIS for 31\,500 s in three visits
between January 31 and February 2, 2000 with the medium resolution
NUV echelle mode (E230M) and a $0.2 \times 0.2$ aperture which
provides a resolution of $\sim$ 30\,000 (FWHM $\simeq 10\,{\rm
km\,s^{-1}}$). We used the HST pipeline data with an additional
correction for inter-order background correction (Rosa, private
communication). The spectrum covers the range between 2279\,\AA\
and $\sim 3080$\,\AA. The coverage at the red end guarantees
overlap with the UVES spectra which extend shortwards to $\sim$
3050\,\AA.

\subsection{VLT/UVES spectroscopy}
 Echelle spectra of HE 0515$-$4414 were obtained during commissioning of
 UVES at the VLT/Kueyen telescope.  The observations were carried out
 under good seeing conditions (0.5 -- 0.8\,arcsec) and a slit width
 of 0.8\,arcsec was used. A summary of the observations and of the
 detectors used is given in the ESO web pages {\tt
 http://www.hq.eso.org/instruments/uves}.

The spectra were extracted using an algorithm that attempts to reduce
 the statistical noise to a minimum. After bias-subtracting and
 flat-fielding of the individual CCD frames, the seeing profiles were
 fitted with a Gaussian in two steps. In a first step the three
 parameters of the Gaussian -- width, amplitude, and offset from the
 previously defined orders -- were unconstrained; in the second step
 only the amplitudes were allowed to vary, with width and offset held
 fixed at values found by a $\kappa\sigma$-clipping fit along the
 dispersion direction to the values obtained in the first step.  Flux
 values were assigned with a variance according to the Poisson
 statistics and the read-out noise, while cosmic-ray shots were
 assigned with infinite variances. Thus, the extraction procedure
 recovers the total count number even at wavelengths where the spatial
 profile is partially modified by cosmic-ray hits.

 The extracted spectra were wavelength calibrated using as reference
  Th-Ar spectra taken  after each science exposure.  All
  wavelength solutions were accurate to better than typically $1/10$
  pixel. The wavelength values were converted to vacuum heliocentric
  values and each spectrum of a given instrumental configuration was
  binned onto a common linear wavelength scale (of typically $0.04$
  \AA\ per pixel). Finally, the reduced spectra were added
  weighting by the inverse of the flux variances.

\subsection{Line profile analysis}

Our analysis was carried out using a multiple line fit procedure to
determine the parameters $\lambda_{\rm c}$ (line center wavelength),
$N$ (column density), and $b$ (line broadening velocity)
for each absorption component. We have written a FORTRAN program based on the
Levenberg-Marquardt algorithm to solve this nonlinear regression problem
(see, e.g., Bevington \& Robinson 1992). We have included additional
parameters describing the local continuum curvature by a low order
Legendre polynomial. A free floating continuum is a prerequisite for an adequate
profile decomposition in the case of complex line ensembles.

To improve the numerical efficiency we have to provide adequate initial parameters.
In some cases the success of the fitting depends on good starting parameters, since
the algorithm tends to converge to the nearest, not necessarily global,
minimum of the chi-square merit function. A first approximation can be found neglecting the
instrumental profile and
converting the flux profile into apparent optical depths using the relation
\begin{equation}
\tau_{\rm a}(\lambda) = \ln [F_{\rm c}(\lambda)/F_{\rm obs}(\lambda)]\,,
\end{equation}
where $F_{\rm c}$ and $F_{\rm obs}$ are the continuum level and
the observed line flux, respectively. If the instrumental
resolution is high compared to the line width, $\tau_{\rm
a}(\lambda)$ will be a good representation of the true optical
depth $\tau(\lambda)$. However, an ill-defined continuum level or
saturation effects may produce large uncertainties. The apparent
optical depth can be automatically fitted with a sum of Gaussians,
each having a variable position, amplitude, and width.  An a
priori line identification is not necessary at this stage of the
analysis.

Having obtained first-guess parameters we proceed with Doppler
profile fitting using artificial test lines with $z=0$ and $f=1$,
where $f$ is the oscillator strength. It can be shown that most
Voigt profiles are well represented by the purely velocity
broadened Doppler core. The size of the fit region depends on the
complexity and extent of the absorption line ensembles. Indeed,
the number of free parameters should be less than 100 to preserve
the numerical efficiency. One specific characteristic of our
technique is the simultaneous continuum normalization which can
reconstruct the true continuum level even in cases, where the
background is hidden by numerous lines. The multi component
profile is the convolution of the intrinsic spectrum and  the
instrumental spread function $P(\Delta \lambda)$:
\begin{equation}
F(\lambda) = P(\Delta\lambda) \otimes \left\{ F_{\rm c}(\lambda) \prod_i
\exp[-\tau_i(\lambda,\lambda_{\rm c},N,b)] \right\}\,.
\end{equation}
If the program fails to converge on a reasonable model, the
parameters can be adjusted by hand. In this way the fit can be
modified to be acceptable by eye and then re-minimized. In some
exceptional cases this procedure is the only chance to free a
converged solution from a local chi-square minimum.

After line identification the parameters of the test lines can be transformed to the actual redshift and
oscillator strength. However, the contribution of unknown profile components can still be considered
using the test line results. A final Voigt profile fit with all identified components includes the
simultaneous multiplet treatment, keeping the redshift, column density, and line width the same
during the chi-square minimization.
The upper limit of the column densities of non-detected lines is estimated assuming a $5\,\sigma$
significance level for the equivalent width.

\section{Absorption systems with \ion{O}{vi} lines} We searched for
\ion{O}{vi} lines associated with known Ly $\alpha$ and Ly $\beta$
absorbers. Therefore, as a starting point, we tried to identify
all Ly $\alpha$ lines. Line identification and the analysis of the
Ly $\alpha$ forest will be presented in some detail in a later
paper. At the resolution of $\sim 30\,000$ (STIS) and $\sim
50\,000$ (UVES), narrow metal lines can usually be distinguished
easily from hydrogen lines. In all Ly $\alpha$ absorption systems
with column densities log N$_{H} \geq$ 13.5 we searched for metal
lines, in particular for \ion{O}{vi}, \ion{C}{iv}, \ion{N}{v},
\ion{Si}{iv}, \ion{C}{iii}, \ion{N}{iii}. In a first step, all
lines within $\pm 200\, {\rm km\,s^{-1}}$ were considered to be
plausibly associated with the Ly $\alpha$/Ly $\beta$ systems.
Within this selection criterium we have found 6 systems with
probable \ion{O}{vi} absorption, listed in Table 1. Due to the
moderate S/N ratio of the STIS spectra (between 10 and 20 per
resolution element) the detection limit of \ion{O}{vi} is
estimated to lie between $\log N = 13.3$ at the lower limit of the
z range and $\log N = 13.1$ near the quasar.

\begin{itemize}
\item $z = 1.385$:
The single \ion{O}{vi} line and the \ion{C}{iv} doublet correspond
to the unsaturated Ly $\alpha$ line at 2899.42\,\AA. Both
\ion{O}{vi} and \ion{C}{iv} are slightly blueshifted (by
$-14\,{\rm km\,s^{-1}}$ and $-17\,{\rm km\,s^{-1}}$, respectively)
relative to Ly $\alpha$ and Ly $\beta$ (Fig.\ 1). The Doppler
parameter of the \ion{C}{iv} doublet of $8.2\,{\rm km\,s^{-1}}$ is
determined reliably with the high-resolution, high S/N UVES
spectra, while \ion{O}{vi}, less certain, yields $13.5\,{\rm
km\,s^{-1}}$. The absorbing cloud at z = 1.385 is a close
neighbour to a strong Ly $\alpha$/Ly $\beta$ system at z = 1.386
($+123\,{\rm km\,s^{-1}}$, $\log N_{\rm H} = 15.1$) with no heavy
element absorption at all. The $z = 1.385$ system appears to
represent the extremely rare case of a highly ionized cloud with
low neutral hydrogen density ($\log N_{\rm H} = 13.9$). According
to the velocity centroids, \ion{C}{iv} and \ion{O}{vi} are
apparently not formed in the same volume. In addition, the
velocity shift between \ion{H}{i} and \ion{C}{iv}/\ion{O}{vi} is
in favour of different phases (volumes). This behaviour is similar
to what Tripp et al.\ (2000) have found in a system at z = 0.22637
in H 1821+643. Because apparently \ion{H}{i}, \ion{C}{iv}, and
\ion{O}{vi} are not formed in the same volume, there are no
empirical constraints on the ionization mechanism (photoionization
versus collisional ionization).\\

\item $z = 1.416$:
The absorber is seen in Ly $\alpha$, Ly $\beta$ and \ion{O}{vi}
1031, while the \ion{O}{vi} 1037 line is blended with Ly
$\varepsilon$ of a strong system at $z = 1.674$. Since again a
velocity shift of $+22\,{\rm km\,s^{-1}}$ is seen between
\ion{O}{vi} 1031 and Ly $\alpha$/Ly $\beta$, this system can only
be considered as marginal.\\

\item $z = 1.602$:
Besides Ly $\alpha$ and Ly $\beta$, only the \ion{O}{vi} doublet
is detected at a velocity of $+18\,{\rm km\,s^{-1}}$ relative to
the hydrogen main component. Notice that in velocity space the
\ion{O}{vi} doublet is located
between two Ly $\alpha$ clouds (cf.\ Table 1, Fig.\ 1).\\

\item $z = 1.674$:
This absorbtion system is seen in Ly $\alpha$ down to Ly
$\varepsilon$ and exhibits a strong \ion{O}{vi} doublet. The
\ion{C}{iv} doublet in our UVES spectra show that it consists of
two components, a narrow ($b \simeq 9\,{\rm km\,s^{-1}}$),
stronger component redshifted by $4.3 \,{\rm km\,s^{-1}}$ relative
to hydrogen, and a broad ($b \simeq 18\,{\rm km\,s^{-1}}$)
component at $-12.8\,{\rm km\,s^{-1}}$. Noteworthy are the broad
wings of Ly $\alpha$ which can be explained only with an
additional extremely broad ($b
> 100\,{\rm km\,s^{-1}}$) unsaturated ($\log\, N_{\rm H} = 14.1$)
component which in velocity nearly coincides with the \ion{O}{vi}
line and the saturated Ly $\alpha$ Doppler core ($\log\, N_{\rm H}
= 15.1$). The broad wing is not seen clearly in the other Lyman
lines, but is still consistent with the lower S/N STIS spectra.
This extremely broad component is probably caused by collapsing
or expanding structures in the intergalactic medium.\\

\item $z = 1.697$:
This is a system with \ion{O}{vi}, \ion{C}{iv}, and \ion{N}{v}
doublets as well as \ion{C}{iii} 977\,\AA\ in combination with an
unsaturated Ly $\alpha$ line. The high resolution UVES profiles of
both \ion{C}{iv} and \ion{N}{v} show two components (Fig.\ 1): a
strong one, nearly unshifted relative to hydrogen, and a weak one
at $\sim -33\,{\rm km\,s^{-1}}$. The \ion{O}{vi} 1037\,\AA\ line
is at $+ 0.4\,{\rm km\,s^{-1}}$ with a possible second component
at $-32.9\,{\rm km\,s^{-1}}$ (\ion{O}{vi} 1031\,\AA\ is blended
with a strong Ly $\alpha$ line at $z = 1.2897$). It seems a likely
supposition that \ion{O}{vi} is being formed in the same volume as
\ion{C}{iv} and \ion{N}{v}. If so, the column density ratios
\ion{O}{vi}/\ion{C}{iv} and \ion{O}{vi}/\ion{N}{v} as well as the
high \ion{O}{vi}/\ion{H}{i} ratio are in favour of photoionization
in the proximity zone of the QSO. Notice that
HE~0515-4414 is one of the most luminous QSOs in the universe.\\

\item $z = 1.736$:
This is an associated system close to the systemic QSO redshift
($z_{\rm em} = 1.73$) which shows only the \ion{O}{vi} doublet.
Both \ion{C}{iv} and \ion{N}{v} are not detected, in spite of the
high S/N of the UVES spectra. Again, the inferred lower limits to
the column density ratios \ion{O}{vi}/\ion{C}{iv} and
\ion{O}{vi}/\ion{N}{v} are consistent only with photoionization.
\end{itemize}

\begin{figure*}[h]
  \includegraphics*{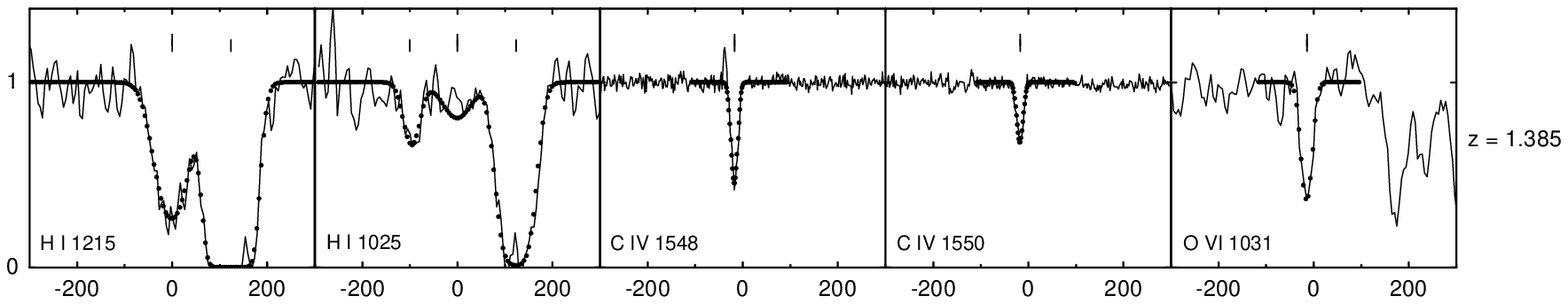}
  \includegraphics*{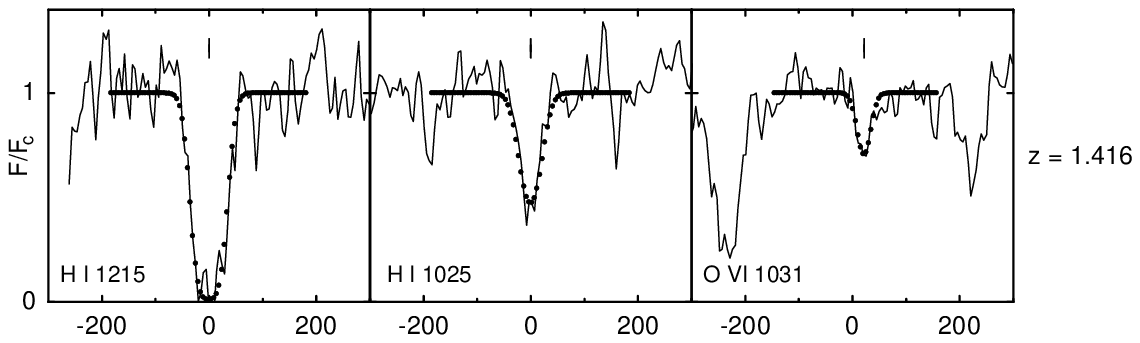}
  \includegraphics*{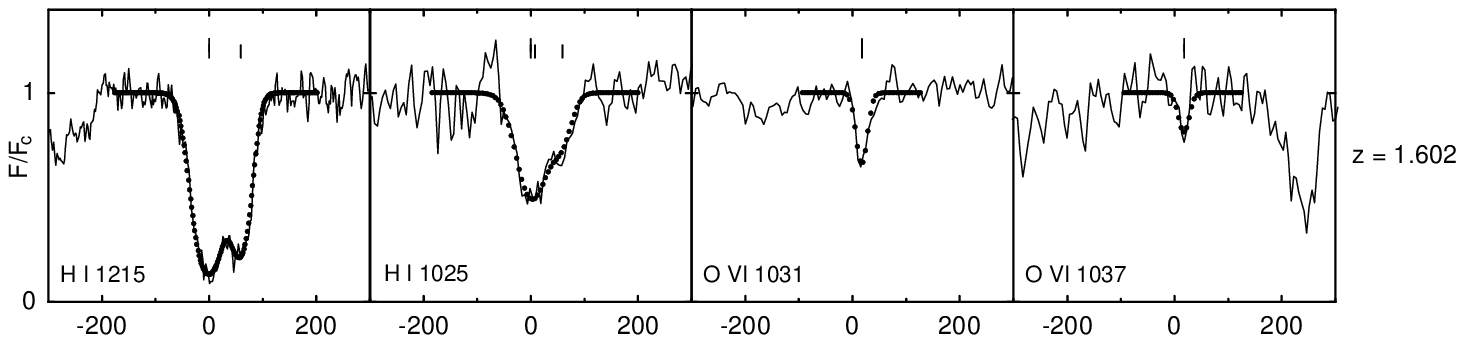}
  \includegraphics*{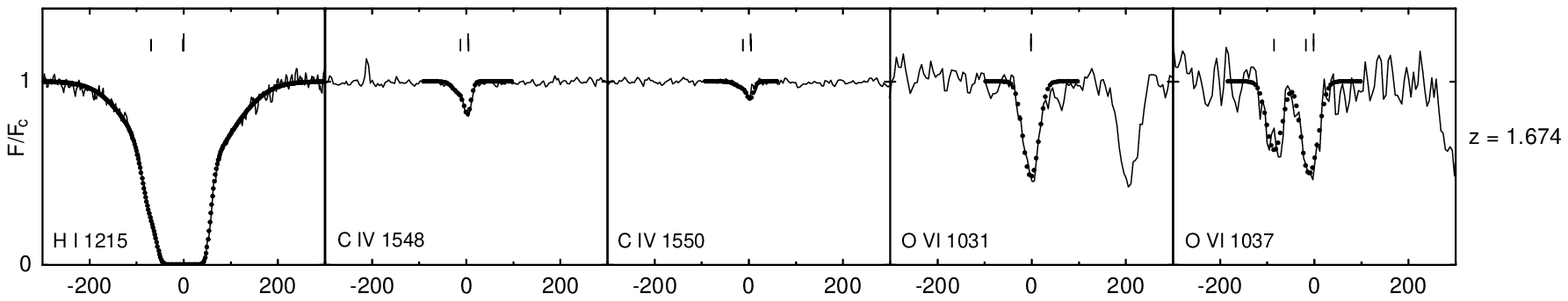}
  \includegraphics*{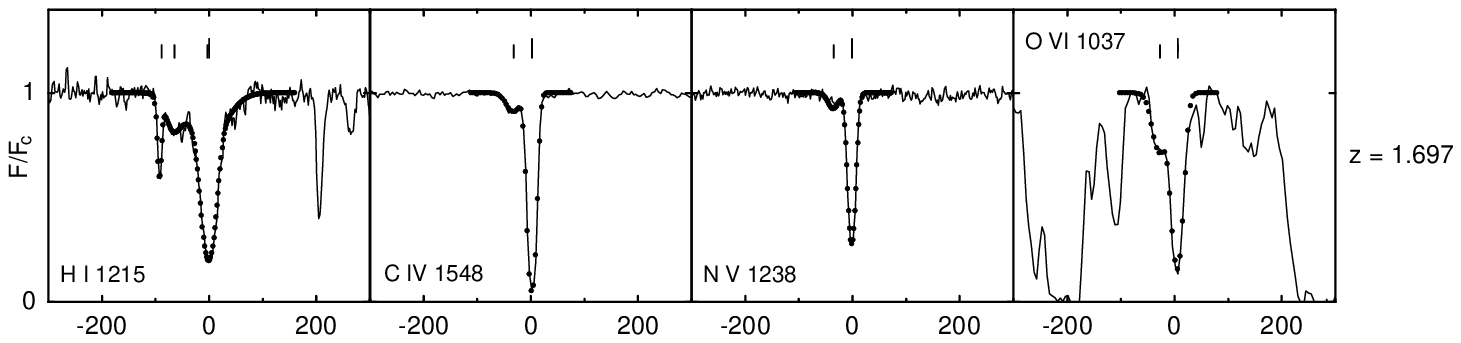}
  \includegraphics*{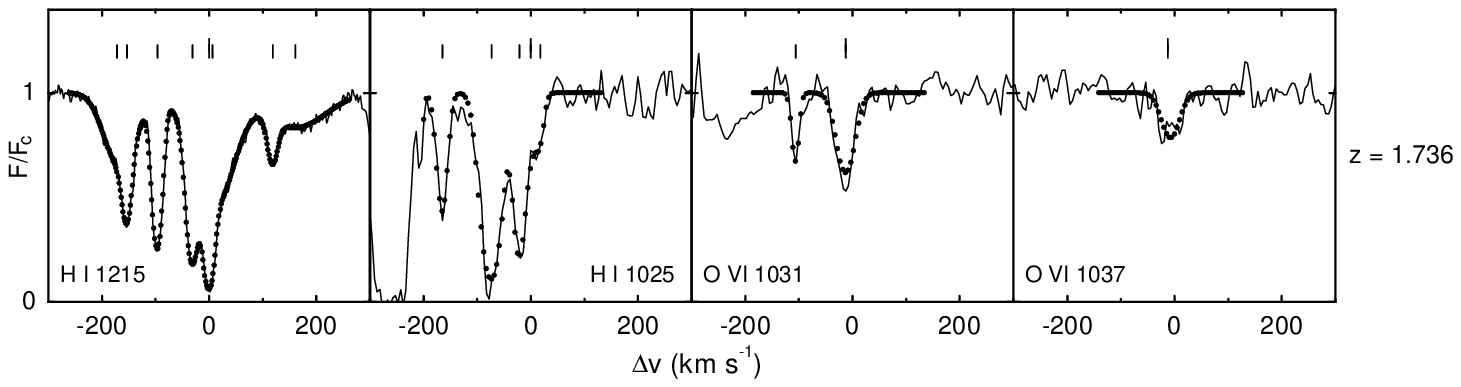}
  \caption{Selected absorption line profiles of systems with \ion{O}{vi} detection. The normalized flux
  is plotted vs. rest-frame velocity of the hydrogen main component. Long tick marks indicate the position of
  the primary lines, while short tick marks indicate additional absorption
  components. It should be noted that some profile ensembles
  contain lines which do not belong to the same absorption system.
  The dotted curves represent our fit models}
  \label{fig1}
\end{figure*}

\begin{table*}[h]
\begin{center}
\caption{\label{absovi1} Absorption line systems with \ion{O}{vi}
detections}
\begin{tabular}{lrcccrcr@{.}l}
\hline {\rule[-2mm]{-1mm}{6mm} Ion} & $\lambda_0$ (\AA) &
$\lambda_{\rm obs}$ (\AA) & $z$ & $\log\,N$ & $b$ (km\,s$^{-1})$&
$W_{\rm rest}$ (m\AA) &
\multicolumn{2}{c}{$\Delta v$ (km\,s$^{-1})$}\\
\hline \\
\underline{z=1.385} & & & & & & & \multicolumn{2}{c}{}\\
\\
\ion{H}{i} & 1025.722 & 2446.383 & $1.385035 \pm 0.000019$ & $13.86 \pm 0.03$ & $40.6 \pm 3.5$ &  50 & 0&0   \\
           & 1025.722 & 2447.391 & $1.386017 \pm 0.000008$ & $15.10 \pm 0.07$ & $32.8 \pm 1.3$ & 309 & +123&4\\
           & 1215.670 & 2899.415 & $1.385035 \pm 0.000019$ & $13.86 \pm 0.03$ & $10.6 \pm 3.5$ & 258 & 0&0   \\
           & 1215.670 & 2900.609 & $1.386017 \pm 0.000008$ & $15.10 \pm 0.07$ & $32.8 \pm 1.3$ & 517 & +123&4\\

\ion{C}{iii}& 977.020 & 2330.005 & $1.384808 \pm 0.000031$ & $12.76 \pm 0.22$ & $7.4 \pm 7.0$  & 28  & $-28$&5 \\

\ion{C}{iv}& 1548.195 & 3692.288 & $1.384899 \pm 0.000001$ & $13.21 \pm 0.01$ & $8.2 \pm 0.2$  & 47  & $-17$&1 \\
           & 1550.770 & 3698.429 & $1.384899 \pm 0.000001$ & $13.21 \pm 0.01$ & $8.2 \pm 0.2$  & 27  & $-17$&1 \\

\ion{Si}{iv}&          &         &           & $< 12.0$  &     &       & \multicolumn{2}{c}{} \\

\ion{N}{v}  &          &         &           & $< 12.8$  &     &       & \multicolumn{2}{c}{} \\

\ion{O}{vi}& 1031.926 & 2461.064 & $1.384923 \pm 0.000012$  &$13.88 \pm 0.06$ &  $13.5 \pm 2.2$ &  65 &  $-14$&1 \\
           & 1037.617 & blend     &           &          &     &       & \multicolumn{2}{c}{} \\
\\
\underline{z=1.416} & & & & & & & \multicolumn{2}{c}{}\\
\\
\ion{H}{i} & 1025.722 & 2478.159 & $1.416014 \pm 0.000011$ & $14.21 \pm 0.05$ & $25.6 \pm 1.6$ & 94 & 0&0\\
           & 1215.670 & 2937.075 & $1.416014 \pm 0.000011$ & $14.21 \pm 0.05$ & $25.6 \pm 1.6$ & 290 & 0&0\\

\ion{C}{iii}& 977.020 & blend & & & & & \multicolumn{2}{c}{} \\

\ion{C}{iv} &          &         &           & $< 11.9$  &     &       & \multicolumn{2}{c}{} \\

\ion{Si}{iv}&          &         &           & $< 11.6$  &     &       & \multicolumn{2}{c}{} \\

\ion{N}{v}  &          &         &           & $< 13.3$  &     &       & \multicolumn{2}{c}{} \\

\ion{O}{vi}& 1031.926 & 2493.327 & $1.416188 \pm 0.000027$ & $13.43 \pm 0.12$ & $14.5 \pm 5.2$ & 30 & +21&6 \\
           & 1037.617 & blend   &         &       &      &       & \multicolumn{2}{c}{}\\
\\
\underline{z=1.602} & & & & & & & \multicolumn{2}{c}{}\\
\\
\ion{H}{i} & 1025.722 & 2668.702 & $1.601779 \pm 0.000016$ & $13.95 \pm 0.03$ & $32.8 \pm 1.9$ &  59 & 0&0   \\
           & 1025.722 & 2669.193 & $1.602288 \pm 0.000016$ & $13.68 \pm 0.04$ & $24.1 \pm 1.8$ &  32 & +58&6 \\
           & 1215.670 & 3162.905 & $1.601779 \pm 0.000016$ & $13.95 \pm 0.03$ & $32.8 \pm 1.9$ & 251 & 0&0   \\
           & 1215.670 & 3163.523 & $1.602288 \pm 0.000016$ & $13.68 \pm 0.04$ & $24.1 \pm 1.8$ & 155 & +58&6 \\

\ion{C}{iii}&          &         &           & $< 12.4$  &     &       & \multicolumn{2}{c}{} \\

\ion{C}{iv} &          &         &           & $< 11.8$  &     &       & \multicolumn{2}{c}{} \\

\ion{Si}{iv}&          &         &           & $< 11.7$  &     &       & \multicolumn{2}{c}{} \\

\ion{N}{v}  &          &         &           & $< 12.5$  &     &       & \multicolumn{2}{c}{} \\

\ion{O}{vi}& 1031.926 & 2685.006 & $1.601937 \pm 0.000017$ & $13.46 \pm 0.07$ & $12.5 \pm 3.6$ & 31 & +18&2 \\
           & 1037.617 & 2699.813 & $1.601937 \pm 0.000017$ & $13.46 \pm 0.07$ & $12.5 \pm 3.6$ & 16 & +18&2 \\
\\
\underline{z=1.674} & & & & & & & \multicolumn{2}{c}{}\\
\\
\ion{H}{i} &  949.743 & 2539.247 & $1.673615 \pm 0.000015$ & $14.07 \pm 0.02$ & $112.7 \pm 3.8$ & 13 & $-1$&4\\
           &  949.743 & 2539.259 & $1.673628 \pm 0.000007$ & $15.07 \pm 0.03$ & $30.9 \pm 0.7$ & 103 & 0&0   \\
           &  972.537 & 2600.189 & $1.673615 \pm 0.000015$ & $14.07 \pm 0.02$ & $112.7 \pm 3.8$ & 27 & $-1$&4\\
           &  972.537 & 2600.201 & $1.673628 \pm 0.000007$ & $15.07 \pm 0.03$ & $30.9 \pm 0.7$ & 173 & 0&0   \\
           & 1025.722 & 2742.386 & $1.673615 \pm 0.000015$ & $14.07 \pm 0.02$ & $112.7 \pm 3.8$ & 82 & $-1$&4\\
           & 1025.722 & 2742.399 & $1.673628 \pm 0.000007$ & $15.07 \pm 0.03$ & $30.9 \pm 0.7$ & 292 & 0&0   \\
           & 1215.670 & 3250.234 & $1.673615 \pm 0.000015$ & $14.07 \pm 0.02$ & $112.7 \pm 3.8$ &466 & $-1$&4\\
           & 1215.670 & 3250.249 & $1.673628 \pm 0.000007$ & $15.07 \pm 0.03$ & $30.9 \pm 0.7$ & 466 & 0&0   \\

\ion{C}{iii}& 977.020 & blend        &           &        &     &       & \multicolumn{2}{c}{} \\

\ion{C}{iv}& 1548.195 & 4139.120 & $1.673514 \pm 0.000112$ & $12.45 \pm 0.37$ & $18.3 \pm 10.1$  & 10  & $-12$&8 \\
           & 1548.195 & 4139.356 & $1.673666 \pm 0.000009$ & $12.59 \pm 0.25$ & $9.1 \pm 2.3$  & 14  & +4&3 \\
           & 1550.770 & 4146.005 & $1.673514 \pm 0.000109$ & $12.45 \pm 0.37$ & $18.3 \pm 10.1$  &  5  & $-12$&8 \\
           & 1550.770 & 4146.241 & $1.673666 \pm 0.000010$ & $12.59 \pm 0.25$ & $9.1 \pm 2.3$  &  7  & +4&3 \\

\ion{Si}{iv}&          &         &           & $< 11.6$  &     &       & \multicolumn{2}{c}{} \\

\ion{N}{v}  &          &         &           & $< 12.2$  &     &       & \multicolumn{2}{c}{} \\

\ion{O}{vi}& 1031.926 & 2758.972 & $1.673614 \pm 0.000013$  &$13.88 \pm 0.04$ &  $20.2 \pm 2.0$ &  74 &  $-1$&2 \\
           & 1037.617 & 2774.187 & $1.673614 \pm 0.000013$  &$13.88 \pm 0.04$ &  $20.2 \pm 2.0$ &  42 &  $-1$&2 \\
\hline
\end{tabular}
\end{center}
\end{table*}

\setcounter{table}{0}

\begin{table*}[h]
\begin{center}
\caption{\label{absovi2} Continued}
\begin{tabular}{lrcccrcr@{.}l}
\hline {\rule[-2mm]{-1mm}{6mm} Ion} &  $\lambda_0$ (\AA) &
$\lambda_{\rm obs}$ (\AA) & $z$ & $\log\,N$ & $b$ (km\,s$^{-1})$&
$W_{\rm rest}$ (m\AA) &
\multicolumn{2}{c}{$\Delta v$ (km\,s$^{-1})$}\\
\hline \\
\underline{z=1.697} & & & & & & \multicolumn{2}{c}{} \\
\\
\ion{H}{i} & 1025.722 & blend   &           &       &     &       & \multicolumn{2}{c}{} \\
           & 1215.670 & 3278.803 & $1.697116 \pm 0.000003$ & $13.48 \pm 0.03$ & $15.9 \pm 0.7$ & 100 &   0&0\\

\ion{C}{iii}& 977.020 & 2635.121 & $1.697101 \pm 0.000013$ & $12.96 \pm 0.08$ & $8.5 \pm 2.3$  & 40  & $-1$&7 \\

\ion{C}{iv}& 1548.195 & 4175.207 & $1.696822 \pm 0.000011$ & $12.61 \pm 0.04$ & $18.6 \pm 2.0$  & 18  & $-32$&7 \\
           & 1548.195 & 4175.682 & $1.697129 \pm 0.000001$ & $13.82 \pm 0.01$ & $8.4 \pm 0.1$  & 122  & +1&5 \\
           & 1550.770 & 4182.151 & $1.696822 \pm 0.000011$ & $12.61 \pm 0.04$ & $18.6 \pm 2.0$  &  9  & $-32$&7 \\
           & 1550.770 & 4182.627 & $1.697129 \pm 0.000001$ & $13.82 \pm 0.01$ & $8.4 \pm 0.1$  &  89  & +1&5 \\

\ion{Si}{iv}& 1393.755& 3759.174 & $1.697156 \pm 0.000016$ & $11.56 \pm 0.09$ & $8.5^*$ &  3  & +4&4 \\
            & 1402.770& 3783.489 & $1.697156 \pm 0.000016$ & $11.56 \pm 0.09$ & $8.5^*$ &  2  & +4&4 \\

\ion{N}{v} & 1238.821 & 3340.882 & $1.696824 \pm 0.000022$ & $12.47 \pm 0.13$ & $10.7 \pm 3.9$  &  6  & $-32$&5 \\
           & 1238.821 & 3341.236 & $1.697109 \pm 0.000002$ & $13.61 \pm 0.01$ & $ 8.4 \pm 0.3$  & 53  & $-0$&8 \\
           & 1242.804 & 3351.623 & $1.696824 \pm 0.000022$ & $12.47 \pm 0.13$ & $10.7 \pm 3.9$  &  3  & $-32$&5 \\
           & 1242.804 & 3351.978 & $1.697109 \pm 0.000002$ & $13.61 \pm 0.01$ & $ 8.4 \pm 0.3$  & 33  & $-0$&8 \\

\ion{O}{vi}& 1031.926 & 2782.919 & $1.696820 \pm 0.000029$  &$13.77 \pm 0.09$ &  $16.7 \pm 4.0$ &  58 &  $-32$&9 \\
           & 1031.926 & 2783.229 & $1.697120 \pm 0.000008$  &$14.40 \pm 0.04$ &  $11.8 \pm 1.3$ & 112 &  +0&4 \\
           & 1037.617 & 2798.266 & $1.696820 \pm 0.000029$  &$13.77 \pm 0.09$ &  $16.7 \pm 4.0$ &  33 &  $-32$&9 \\
           & 1037.617 & 2798.577 & $1.697120 \pm 0.000008$  &$14.40 \pm 0.04$ &  $11.8 \pm 1.3$ &  85 &  +0&4 \\
\\
\underline{z=1.736} & & & & & & & \multicolumn{2}{c}{}\\
\\
\ion{H}{i} & 1025.722 & 2806.274 & $1.735901 \pm 0.000003$ & $13.50 \pm 0.02$ & $10.9 \pm 0.5$ &  21 & 0&0   \\
           & 1215.670 & 3325.952 & $1.735901 \pm 0.000003$ & $13.50 \pm 0.02$ & $10.9 \pm 0.5$ &  86 & 0&0   \\

\ion{C}{iii}&          &         &           & $< 12.4$  &     &       & \multicolumn{2}{c}{} \\

\ion{C}{iv} &          &         &           & $< 11.7$  &     &       & \multicolumn{2}{c}{} \\

\ion{Si}{iv}&          &         &           & $< 11.6$  &     &       & \multicolumn{2}{c}{} \\

\ion{N}{v}  &          &         &           & $< 12.3$  &     &       & \multicolumn{2}{c}{} \\

\ion{O}{vi}& 1031.926 & 2823.133 & $1.735790 \pm 0.000017$ & $13.73 \pm 0.05$ & $22.0 \pm 2.8$ & 57 & $-12$&2 \\
           & 1037.617 & 2838.701 & $1.735790 \pm 0.000017$ & $13.73 \pm 0.05$ & $22.0 \pm 2.8$ & 32 & $-12$&2 \\
\hline
\end{tabular}
\begin{list}{}{}
  \item[$^*$] The Doppler parameter has been fixed to improve the
  goodness-of-fit
\end{list}
\end{center}
\end{table*}

\section{The cosmological mass-density of the \ion{O}{vi} phase}
With the present STIS spectra of HE~0515-4414 the redshift range
$1.21 \leq z \leq 1.73$ has been covered for the first time at
sufficiently high resolution to undertake a sensitive search for
\ion{O}{vi} absorbers. We have detected 6 \ion{O}{vi} systems. Two
of them (z = 1.697, 1.736) are either associated with the QSO or
in the proximity zone of the extremely luminous QSO. The system $z
= 1.416$ is marginal, since only the 1031\,\AA\ line is detected.
Including the latter, we have 4 detections in the range z = 1.21
to 1.67 which yield a number density of \ion{O}{vi} absorbers with
$W_{\rm rest} \geq$ 25\,m\AA\ of ${\rm d}N/{\rm d}z \leq 10$.
Compared with the findings by Tripp et al.\ (2000) of ${\rm
d}N/{\rm d}z = 48$ at $\bar{z} \simeq 0.21$, the number density at
$\bar{z} \simeq 1.44$ is roughly a factor of 5 lower. Tripp et
al.\ (2000) compared their finding of a high number density of
weak \ion{O}{vi} absorbers ($W_{\rm rest} \geq$ 30\, m\AA) in
H~1821+643 and PG~0953+415 with other classes of absorbers and
found that the weak \ion{O}{vi} number density is more comparable
to that of the low $z$ weak Ly $\alpha$ absorbers -- which have
${\rm d}N/{\rm d}z \approx 100$ for $W_{\rm rest} \geq$ 50\,m\AA\
-- than to other types of metal absorbers like \ion{Mg}{ii}. In
HE~0515-4414 we have at least 42 Ly $\alpha$ systems (the exact
number being unknown due to the line blending problem) with
$W_{\rm rest} \geq $ 50\,m\AA\ in the range $1.21 \leq z \leq
1.67$ which yields roughly ${\rm d}N/{\rm d}z = 90$ \footnote{A
more detailed discussion of the Ly $\alpha$ forest in HE~0515-4414
is in progress}. Among these, roughly half of them are strong,
saturated Ly $\alpha$ lines with a detected Ly $\beta$ line.
Again, while our STIS spectrum of HE~0515-4414 confirms the number
density of Ly $\alpha$ absorbers found previously (see Weymann et
al.\ 1998), the number of \ion{O}{vi} absorbers with $W_{\rm rest}
\geq 25$\,m\AA\ is lower than the number of Ly $\alpha$ absorbers
with $W_{\rm rest} \geq 50$\,m\AA\ by a factor of 10. It is
noteworthy that, except the $z = 1.674$ system, \ion{O}{vi} is
detected in lower column density Ly $\alpha$ absorbers (log
$N_{\rm H} \leq 14$). Following the calculations by Tripp et al.\
(2000) and earlier work by Storrie-Lombardi et al.\ (1996) and
Burles \& Tytler (1996), the mean cosmological mass-density of
\ion{O}{vi} absorbers can be written in units of the critical
density $\rho_{\rm c}$ as
\begin{equation}
\Omega_{\rm b} (\ion{O}{vi}) = \frac{\mu\,m_{\rm
H}\,H_{0}}{\rho_{\rm c}\,c\,f(\ion{O}{vi})}\, \left(\frac{\rm
O}{\rm H}\right)^{-1}_{\ion{O}{vi}} \frac{\sum_i N_i ({\rm
\ion{O}{vi}})} {\Delta X},
\end{equation}
where [O/H] is the assumed oxygen abundance in the \ion{O}{vi}
absorbing gas, $f$(\ion{O}{vi}) is the fraction of oxygen in
\ion{O}{vi}, $\sum_{i} N_i(\ion{O}{vi})$ is the total \ion{O}{vi}
column density from all absorbers, and $\Delta X$ is the
absorption distance (Bahcall \& Peebles 1969).

Over the redshift interval $z = 1.21$ to $z = 1.67$ we have
$\Delta X = 0.72$ for $q_{0} = 1/2$. $\sum_i N_i ({\rm
\ion{O}{vi}})$ is $2.1 \times 10^{14}\,{\rm cm^{-2}}$ (Table 1).
Assuming $f(\ion{O}{vi}) = 0.2$, following Tripp et al.\ (2000)
and Tripp \& Savage (2000), which is close to the maximum for both
collisional ionization and photoionization, we obtain a lower
limit $\Omega_{\rm b}$(\ion{O}{vi}) $\geq 3 \times 10^{-5}$\,
[(O/H)/(O/H)$_{\odot}]^{-1}\,h^{-1}_{75}$. The only reliably
measured heavy element abundances at $\bar{z} = 1.4$ are from
DLAs. Typically the metal abundance (e.g.\ from Zn) is 1/10 solar
(Pettini et al.\ 1999, Vladilo et al.\ 2000). There is, however,
no guarantee that these abundances apply also to the \ion{O}{vi}
absorbers among the low column density systems. Assuming 1/10
solar for the oxygen abundance, we have $\Omega_{\rm b} ({\rm
\ion{O}{vi}}) \geq 3\times 10^{-4}\, h^{-1}_{75}$. With the same
assumptions Tripp et al.\ (2000) derived a value $\geq 4\times
10^{-3}\, h^{-1}_{75}$. Using a somewhat different formalism for
the derivation of $\Omega_{b}(\ion{O}{vi})$, namely Eq. (6) from
Tripp \& Savage (2000), we get with the same assumptions
$\Omega_{\rm b}$(\ion{O}{vi}) $\geq 1.5 \times 10^{-4}\,
h^{-1}_{75}$. Both from the number counts of the \ion{O}{vi}
systems and the estimate of the mean \ion{O}{vi} density the
unavoidable conclusion seems to be that at $\bar{z} = 1.5$, the
baryon content of the \ion{O}{vi} phase contains a factor of $\geq
10$ less material than at $\bar{z}= 0.21$.

\section{Conclusions}
Our results on \ion{O}{vi} absorbing clouds in HE~0515-4414 can be
summarized as follows:

\begin{itemize}
\item Intervening \ion{O}{vi} systems per unit redshift appear to be less frequent
by a factor of $\sim$ 5 at $\bar{z} = 1.5$ compared to the local
universe $\bar{z}= 0.21$.
\item According to the observed velocities, the \ion{O}{vi} lines are not formed in
the same volume as the \ion{H}{i} and \ion{C}{iv} absorbing
material. This conclusion is supported by the derived Doppler
parameters (see Table 1). An identical gas phase would require
$b_{\rm \ion{O}{vi}} \le b_{\rm \ion{C}{iv}}$ in contradiction to
the observations.
\item Consequently, even narrow related \ion{C}{iv} lines which would eliminate
collisional ionization as mechanism for \ion{C}{iv} production,
cannot rule out collisional ionization for \ion{O}{vi} from
observations. For the same reason, \ion{O}{vi}/\ion{C}{iv} column
density ratios cannot be used for arguing in favour of or against
photoionization/collisional ionization.
\end{itemize}

The occurrence of an extremely broad component superimposed on the
"normal" Doppler profile as observed in the $z = 1.674$ system is
a rare Ly $\alpha$ profile type. In fact, we have never seen such
a profile combination. In the context of modern interpretations of
the Ly $\alpha$ forest as caused by a gradually varying density
field characterized by a network of filaments and sheets (e.g.\ Bi
\& Davidsen), a multi-component Voigt profile fitting is
artificial and without a physical meaning anyway. The observed Ly
$\alpha$ profile at $z = 1.674$ could be easily modelled by an
overdense structure with inflow or outflow velocities of the order
of $100\,{\rm km\,s^{-1}}$. We abstain from such an exercise,
since the line profile decomposition would not lead to a unique
solution. Remarkable is the coincidence with a strong \ion{O}{vi}
doublet at the same velocity.

Our finding, that the \ion{O}{vi} phase at $\bar{z}$ = 1.5§
contains a factor of $\geq 10$ less material than at z = 0.21,
provided the \ion{O}{vi}/O ratio and the oxygen abundance are
similar, appears to be inconsistent with the simulations of Dav\'e
et al.\ (2001) who predict an increase of the mass-fraction of
baryons in the warm-hot phase of the IGM by at most a factor of 4
between $z = 1.5$ and 0.2. An increase in the mean oxygen
abundance in the low density IGM by a factor of $\sim 3$ over the
same redshift range would restore consistency with the theoretical
predictions. However, at present we do not see a possibility to
test this hypothesis. Furthermore, as long as we do not understand
the ionization to \ion{O}{vi} quantitatively, the fractional
ionization \ion{O}{vi}/O might vary between $z = 1.5$ and 0.2.
Finally, our result is still debatable due to small number
statistics. More lines of sight, both at low and intermediate
redshift, have to be probed.

\acknowledgements{This work has been supported by the
Verbundforschung of the BMBF/DLR under Grant No. 50 OR 99111. S.L.
acknowledges financial support by FONDECYT grant N$^{\rm o}
3\,000\,001$ and by the Deutsche Zentralstelle f\"ur
Arbeitsvermittlung.}


\begin{thebibliography}{}
\bibitem[1969]{bah:}Bahcall J.N., Peebles P.J. E., 1969, ApJ 156, L7
\bibitem[1992]{bev:}Bevington P.R., Robinson D.K., 1992, Data Reduction and Error
  Analysis for the Physical Sciences, McGraw-Hill, New York
\bibitem[1997]{bi:}Bi H., Davidsen A.F., 1997, ApJ 479, 523
\bibitem[1996]{bur:}Burles S., Tytler D., 1996, ApJ 460, 584
\bibitem[1999]{cen:}Cen R., Ostriker J.P., 1999, ApJ 519, L109
\bibitem[2001]{dav:}Dav\'e R., Cen R., Ostriker J.P. et al., 2001, ApJ
552, 473
\bibitem[2000]{hea:}Heap S., Williger G.M., Smette A. et al., 2000,
ApJ 534, 69
\bibitem[1999]{pet:}Pettini M., Ellison S.L., Steidel C.L., Bowen D.V., 1999,
ApJ 510, 576
\bibitem[1997]{rei1:}Reimers D., K\"ohler S., Wisotzki L. et al., 1997,
A\&A 327, 890
\bibitem[1998]{rei2:}Reimers D., Hagen H.-J., Rodriguez-Pascual P.,
Wisotzki L., 1998, A \& A 334, 96
\bibitem[1998]{sav:}Savage B.D., Tripp T.M., Lu L., 1998, AJ 115, 436
\bibitem[2000]{sch:}Schaye J., Rauch M., Sargent W.L.W., Kim
T.-S., 2000, ApJ 541, L1
\bibitem[1996]{sto:}Storrie-Lombardi L.J., McMahon R.G., Irwin M.J., 1996,
MNRAS 283, L79
\bibitem[2000]{tri1:}Tripp T.M., Savage B.D., 2000, ApJ 542, 42
\bibitem[2000]{tri2:}Tripp T.M., Savage B.D. Jenkins E.B., 2000, ApJ 534, L1
\bibitem[2000]{vla:}Vladilo G., Bonifacio P., Centurion M., Molaro P., 2000,
ApJ 543, 24
\bibitem[1998]{wey:}Weymann R., Januzzi B.T., Lu L. et al., 1998, ApJ
506, 1

\end{thebibliography}
\end{document}